\documentclass[prb,showpacs,twocolumn,amsmath,amssymb,groupaddress,superscriptaddress]{revtex4}
\usepackage{graphicx}% Include figure files
\usepackage{dcolumn}% Align table columns on decimal point
\usepackage{bm}% bold math

\begin{document}

\preprint{APS/PRB -Girard}

\title{Complex room temperature ferrimagnetism induced by zigzag oxygen-vacancy stripes in Sr$_3$YCo$_4$O$_{10.72}$}

\author{D.D. Khalyavin}
\affiliation{ISIS facility, Rutherford Appleton Laboratory-STFC,
Chilton, Didcot, Oxfordshire, OX11 0QX, UK. }
\author{L.C. Chapon}
\affiliation{ISIS facility, Rutherford Appleton Laboratory-STFC,
Chilton, Didcot, Oxfordshire, OX11 0QX, UK. }
\author{E. Suard}
\affiliation{Institut Laue-Langevin, 6 Jules Horowitz, BP156, 38042 Grenoble Cedex 9, France. }
\author{J.E. Parker}
\affiliation{Diamond Light Source, Harwell Science and Innovation Campus, Didcot, Oxfordshire
OX11 0DE, UK. }
\author{S.P. Thompson}
\affiliation{Diamond Light Source, Harwell Science and Innovation Campus, Didcot, Oxfordshire
OX11 0DE, UK. }
\author{A.A. Yaremchenko}
\affiliation{Department of Ceramics and Glass Engineering, CICECO, University of Aveiro, 3810-193 Aveiro, Portugal. }
\author{V.V. Kharton}
\affiliation{Department of Ceramics and Glass Engineering, CICECO, University of Aveiro, 3810-193 Aveiro, Portugal. }
\date{\today}% It is always \today, today,

\begin{abstract}
The high temperature ferromagnetism in Sr$_3$YCo$_4$O$_{10+\delta}$ perovskite, whose origin has been the subject of a considerable debate, has been studied by neutron powder diffraction and synchrotron X-ray diffraction measurements. Oxygen vacancy ordering creates a complex pattern of zigzag stripes in the oxygen-deficient CoO$_{4+\delta}$ layers, where the Co ions are found in three distinct coordinations. The symmetry of this unprecedented structural modulation, in conjunction with the existence of different Co spin states, provide a straightforward explanation for the appearance of ferrimagnetism. A model for the magnetic structure compatible with these structural features is proposed, based on the refinement of powder neutron data. The macroscopic moment as a function of temperature that can be calculated from the values of the ordered spins extracted from refinements, is in excellent agreement with bulk magnetization. Unlike previous models, a collinear G-type magnetic structure with uncompensated moments due to distinct spin-states of Co imposed by different coordination is found.   
\end{abstract}

\maketitle

\begin{figure}[b]
\includegraphics[scale=1.1]{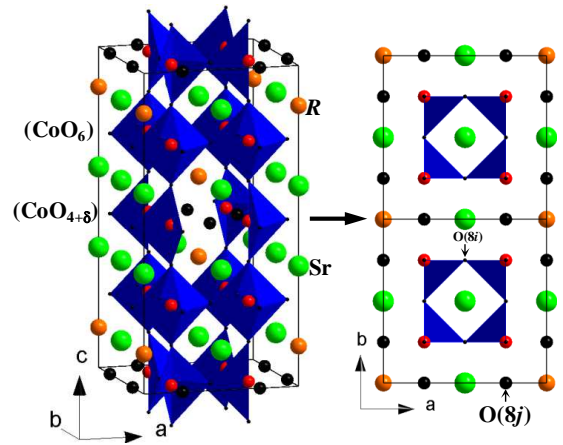}
\caption{(Color online) Schematic representation of the tetragonal $I4/mmm$ ($2a_p\times 2a_p\times 4a_p$) crystal structure of Sr$_3R$Co$_4$O$_{10+\delta}$ perovskites as alternation of the oxygen-occupied (CoO$_6$) and oxygen-deficient (CoO$_{4+\delta}$) layers stacked along the $c$ axis (left). Expanded view of the CoO$_{4+\delta}$ layer with partially occupied $8j$ oxygen position (right).}
\label{fig:par}
\end{figure}

\indent The rich physical properties of cobalt oxides compared to other \emph{3d} transition metal oxides originate in the various electronic states of cobalt ions. Probably the first and most famous example is LaCoO$_3$ perovskite that undergoes a diamagnetic to paramagnetic transition on warming. The phenomenon, interpreted by Goodenough\cite{ISI:A1958WH89000022} as a thermally activated crossover of Co$^{3+}$ from a low-spin to a high-spin state, is due to a subtle balance between interatomic exchange energy and crystal field splitting. Nowadays many complex cobalt-oxides with fascinating electrical and magnetic properties  are known, displaying superconductivity, near room-temperature giant magnetoresistance, high ionic/electronic conductivity and large thermoelectric power, making them attractive and technologically relevant.\cite{ISI:000273727200002} Practically in all cases the Co electronic configuration, primarily determined by the crystalline electric field created by first-neighbour oxygen ions, plays a central role in the underlying physics. The local Co environment is therefore a great lever to tune the electric properties and consequently the magnetic properties (spin-state) of such systems. \\ 
\indent Recently, much attention has been devoted to the oxygen-deficient perovskites Sr$_3R$Co$_4$O$_{10+\delta}$ ($R$=rare earth or Y)\cite{ISI:000184606500027,ISI:000186050800010,ISI:000247624800009,ISI:000267395800029} in particular systems with oxygen content $0.5<\delta <1$ which display unconventional ferromagnetism with the highest critical temperature (T$_m\sim 360$K) among the known cobalt-perovskites. The basic crystal structure of Sr$_3R$Co$_4$O$_{10}$ of tetragonal $I4/mmm$ symmetry, involves both cation ordering (Sr/$R$) and oxygen vacancy ordering with a $2a_p\times 2a_p\times 4a_p$ superstructure with respect to the  pseudo-cubic perovskite unit-cell (Fig. \ref{fig:par}). The latter ordering produces an alternate stacking of \emph{oxygen-rich} octahedral (CoO$_6$) layers and \emph{oxygen-deficient} tetrahedral (CoO$_4$) layers along the $c$-axis. 
For compositions with $0.5<\delta <1$, an additional superstructure has been identified ($2\sqrt{2}a_p\times 2\sqrt{2}a_p\times 4a_p$) and attributed to the ordering of the extra oxygen ions in the CoO$_{4+\delta}$ layers.\cite{ISI:000184606500027,ISI:000186050800010,ISI:000247624800009,ISI:000267395800029,ISI:000249186100007} It has been shown\cite{ISI:000267395800029,ISI:000255043200783,ISI:0} that the formation of such superstructure is key to the appearance of ferromagnetism; however a clear understanding of its microscopic origin is still missing in spite of several neutron diffraction studies. \cite{ISI:000222078700005,ISI:000221278100072,ISI:000250619800065,ISI:000268617500060,ISI:000269956600027,ISI:000268615700046} The main difficulty lies in the simultaneous determination of the oxygen ordering superstructure and the magnetic structure from diffraction data, required to built a comprehensive picture.\\ 
\indent In the present letter, we report a model for the magnetic structure and oxygen vacancy superstructure of Sr$_3$YCo$_4$O$_{10.72}$ obtained by neutron diffraction and symmetry considerations. The magnetic configuration compatible with the superstructure formed by oxygen vacancy ordering, explains the origin of the high temperature ferromagnetism. We show that the oxygen vacancies create unconventional zigzag stripes in the CoO$_{4+\delta}$ layers with three distinct Co environments. Although all nearest-neighbour exchange interactions are strongly antiferromagnetic, the 
symmetry and presence of three inequivalent magnetic sites in the oxygen-deficient layers result in a net spontaneous moment. The magnitude of this ferromagnetic component calculated from the magnetic configuration as a function of temperature is in remarkable quantitative agreement with the magnetization.\\  
\indent A powder sample of Sr$_3$YCo$_4$O$_{10+\delta}$ was synthesized by solid-state reaction as described previously.\cite{ISI:000184606500027} The oxygen content determined by reduction in a 10\% H$_2$ - 90\% N$_2$ mixture using a Setaram Setsys 16/18 thermogravimetric equipment was found to be 10.72(3). Neutron diffraction data were collected at the ILL (Grenoble, France) using the high-resolution D2B (10K$<$T$<$575K) and high intensity D1B (10K$<$T$<$330K) two-axis powder diffractometers. Synchrotron X-ray diffraction data were collected at the Diamond light source (RAL, UK) on the high resolution powder beamline I11\cite{ISI:000265509000007} (10K$<$T$<$600K). The program FullProf\cite{ISI:A1993ME99200007} was employed for Rietveld refinements and group-theoretical calculations were done with the aid of the ISOTROPY software.\cite{ISOTROPY} \\
\begin{figure}[t]
\includegraphics[scale=0.8]{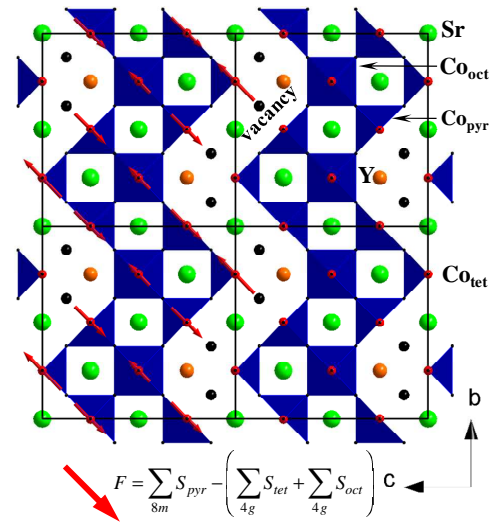}
\caption{(Color online) Schematic representation of the CoO$_{4+\delta}$ layer with oxygen vacancy superstructure associated with the $X^-_4$ irrep. Magnetic moments localized on Co sites with six-fold ($4g$), five-fold ($8m$) and four-fold ($4g$) coordination are shown as arrows of different length. For clarity they are shown in the ($bc$) plane but in the actual refinement the moments were perpendicular to this plane.}
\label{fig:str}
\end{figure}
\indent Analysis of the synchrotron X-ray diffraction data revealed three structural phase transitions at T$_{S1}\sim $ 550K, T$_{S2}\sim $ 330K and T$_{S3}\sim $ 280K. Above T$_{S1}$, the symmetry is tetragonal $I4/mmm$ with a $2a_p\times 2a_p\times 4a_p$ supercell as originally proposed by Istomin et al.\cite{ISI:000186050800010} for Sr$_{0.7}$Y$_{0.3}$CoO$_{2.63}$ and by Withers, James and Goossens\cite{ISI:000184606500027} for $R_{1/3}$Sr$_{2/3}$CoO$_{3-\gamma }$ ($R$=Y, Ho and Dy). The model implies fully occupied $8i$ and partially occupied $8j$ oxygen sites in the oxygen deficient CoO$_{4+\delta }$ layers (Fig. \ref{fig:par}). Below T$_{S1}$, a set of new reflections associated with the $X$(\textbf{\textit{k}}=\textbf{\textit{a*}}/2+\textbf{\textit{b*}}/2) reciprocal point appears, as already reported for the analogue \cite{ISI:000247624800009} Sr$_{3.12}$Er$_{0.88}$Co$_4$O$_{10.5}$. In agreement with previous studies,\cite{ISI:000184606500027,ISI:000186050800010,ISI:000247624800009,ISI:000267395800029,ISI:000249186100007}, we found that the primary distortion involves oxygen vacancy ordering in the CoO$_{4+\delta}$ layers on the partially occupied $8j$ Wyckoff site. 
\begin{table}[b]
\caption{Isotropy subgroups associated with order-disorder modes on the $8j$ Wyckoff position, along a single-arm direction of the irreducible representations of the $I4/mmm$ space group, conjugated with the $X$ point of symmetry.}% title of Table
\centering % used for centering table
\begin{tabular*}{0.48\textwidth}{@{\extracolsep{\fill}} l c c c } % centered columns (4 columns)
\hline\hline %inserts double horizontal lines
Subgroup & Irrep & Lattice vectors & Origin \\ [0.5ex] % inserts table
%heading
\hline % inserts single horizontal line
$Cmmm$ & $X^+_1$ & (0,0,1),(1,1,0),(-1,1,0) & (0,0,0) \\ % inserting body of the table
$Cmca$ & $X^+_2$ & (0,0,1),(1,1,0),(-1,1,0) & (0,0,0) \\
$Cmcm$ & $X^-_3$ & (0,0,1),(1,1,0),(-1,1,0) & (0,1/2,0) \\
$Cmma$ & $X^-_4$ & (0,0,1),(1,1,0),(-1,1,0) & (0,1/2,0) \\ [1ex] % [1ex] adds vertical space
\hline\hline %inserts single line 
\end{tabular*}
\label{table:subgr} % is used to refer this table in the text
\end{table}
Group-theoretical arguments dictate that the symmetry of the superstructure should be compatible with one of the isotropy subgroups given in Table \ref{table:subgr}. Extinction conditions and structural refinements are consistent with a unique subgroup, $Cmma$ associated with the $X^-_4$ irreducible representation (irrep) of the $I4/mmm$ space group. This symmetry coincides with that deduced by Withers, James and Goossens in their original work\cite{ISI:000184606500027} based on electron diffraction and used recently by James et al.\cite{ISI:000249186100007} to refine the room temperature diffraction data of Sr$_{0.8}R_{0.2}$CoO$_{2.5+\delta /4}$.
\begin{figure}[t]
\includegraphics[scale=1.01]{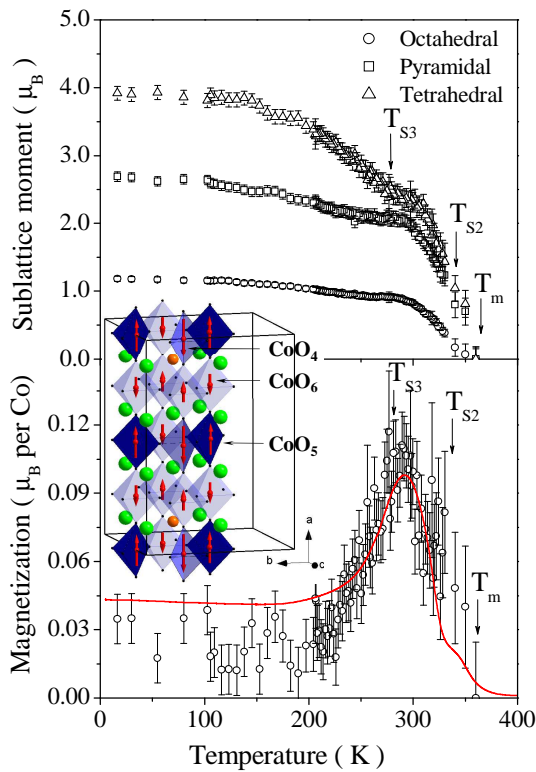}
\caption{(Color online) Magnetic order parameters for octahedral, pyramidal and tetrahedral sublattices as a function of temperature (top). SQUID magnetization measured at H=0.5T at warming after cooling in this field (solid line). Open circles represent net magnetic moment obtained from the neutron diffraction data as a difference between spin values in pyramidal and summarized octahedral and tetrahedral sublatttices (bottom). Inset shows a part of the magnetic unit cell.}
\label{fig:mag}
\end{figure} 
\begin{figure}[t]
\includegraphics[scale=0.9]{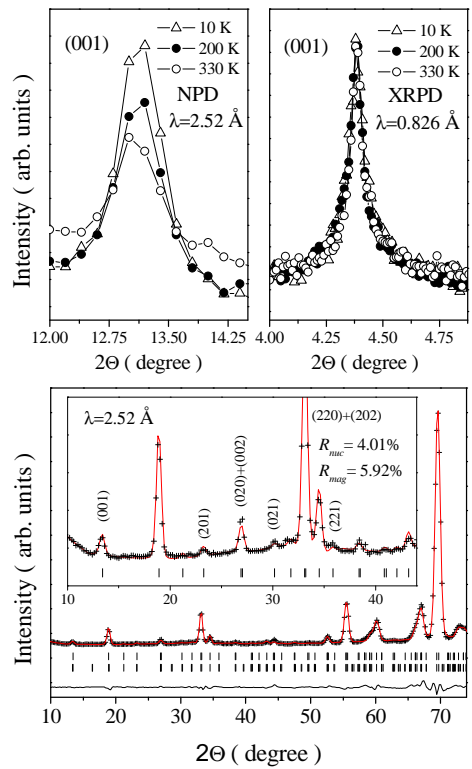}
\caption{(Color online) A part of the neutron (left) and X-ray (right) diffraction patterns in a vicinity of the (001) reflection collected at different temperatures (top). An example of the D1B neutron diffraction pattern refined in the model described in the text (bottom). The cross symbols and (red) solid line represent the experimental and calculated intensities, respectively, and the line below is the difference between them. Tick marks indicate the positions of Bragg peaks for the nuclear scattering ($Cmma$ space group, top line) and magnetic scattering (\textbf{\textit{k}}=0 propagation vector, bottom line).}
\label{fig:diff}
\end{figure}
The corresponding superstructure is shown in Figure \ref{fig:str}. Note that the $c$-axis of the tetragonal structure becomes $a$-axis in the $Cmma$ setting (see Table \ref{table:subgr}). For clarity all atoms are displayed in their highly symmetric positions. The oxygen vacancies create zigzag stripes in the ($bc$)-plane providing three different Co coordinations in the CoO$_{4+\delta}$ layers, namely octahedral (position $4g$), square pyramidal ($8m$) and tetrahedral ($4g$). The presence of these distinctly coordinated Co ions is the crucial ingredient to successfully model the low temperature magnetic scattering and adequately interpret the magnetic properties of this material. There is no evidence from the diffraction data that the vacancy ordering pattern changes with temperature as the critical temperatures of subsequent transitions T$_{S2}$ and T$_{S3}$ are too low for a superstructure reconstruction which requires atomic diffusion on distances comparable to $a_p$. Instead, these transitions are likely to be displacive. At T$_{S2}$, one identifies a doubling of the cell along the $c$ axis with respect to the $Cmma$ space group ($2\sqrt{2}a_p\times 4a_p\times 4\sqrt{2}a_p$ supercell). The appropriate structural models with possible monoclinic symmetry involve more than sixty independent atoms in the unit-cell and cannot be properly refined from the powder data. \\
\indent In agreement with magnetization measurements (Fig. \ref{fig:mag} bottom), the magnetic transition in Sr$_3$YCo$_4$O$_{10.72}$ takes place at T$_m\sim$ 360K. The transition is characterized by the onset of a spontaneous moment whose magnitude reaches a maximum at 290K and then gradually decreases becoming temperature independent below 200K. The magnetic phase transition takes place at a notably higher temperature (T$_m\sim$ 360K) than T$_{S2}$. This observation contradicts previous reports \cite{ISI:000247624800009} of a simultaneous structural and magnetic transition which suggested that ferromagnetism was induced by the change in crystal structure. In the neutron data, additional intensity on top of several nuclear reflections is observed in the magnetically ordered phase, indicating a \textbf{\textit{k}}=0 propagation vector. The largest contribution appears on the 220 reflection  (d-spacing $\sim$4.4 $\AA$) in agreement with previous neutron diffraction studies.\cite{ISI:000222078700005,ISI:000221278100072,ISI:000250619800065,ISI:000268617500060,ISI:000269956600027,ISI:000268615700046}\\
\indent The common interpretation of the magnetic neutron diffraction pattern involves the so called G-type magnetic ordering, where neighbour spins aligned along the $c$ axis of the $I4/mmm$ space group have opposite directions. In early reports, the magnetic structure was refined assuming equal values for all the spins in a unit cell.\cite{ISI:000222078700005,ISI:000221278100072,ISI:000250619800065} Later, it was established  that the Co ions have different spin values 
in the CoO$_6$ and CoO$_{4+\delta }$ layers.\cite{ISI:000268617500060,ISI:000269956600027,ISI:000268615700046} In all these models, ferromagnetism was inferred from spin-canting. This explanation is incompatible with our diffraction data: specifically, the appearance of magnetic intensity on top of the nuclear (001) reflection (Fig. \ref{fig:diff} top) below T$_m$, whose magnetic origin is confirmed by the absence of a corresponding intensity in the X-ray data, directly indicates instead the presence of non-equivalent magnetic moments in the CoO$_{4+\delta }$ layers. In other words, the magnetic arrangement corresponds to a simple collinear G-type antiferromagnet (no canting) but with a non-compensated moment due to the inequivalent spin states imposed by the different Co environments. Accordingly, the refinements of the neutron diffraction data were performed using a model with three independent parameters for the Co spin values in octahedral, pyramidal and tetrahedral coordination. In spite of the simplification, since there are actually six independent Co positions even in the $Cmma$ structure, the model works very well and describes correctly all features of the magnetic scattering pattern (Fig. \ref{fig:diff} bottom). The nuclear scattering was satisfactorily modeled in the orthorhombic $Cmma$ space group. Refined occupancies for the oxygen positions in CoO$_{4+\delta}$ layers yield an average oxygen content 10.81(3) to be sightly higher than from the TGA data. At T=10K, the magnetic moment magnitudes for the four-, five- and six-fold coordinated Co are 4.0(1), 2.7(1) and 1.16(5) $\mu _B$, respectively. A lower spin-state is found for the more coordinated ions. This is in agreement with a recent study showing a certain instability for octahedrally coordinated Co in these materials, for which the application of a moderate pressure ($\sim$2GPa) induces a switching to a non-magnetic low-spin state. \cite{ISI:000264768600087}\\
\indent The variation of the moments as a function of temperature has been extracted from all Rietveld refinements (Fig. \ref{fig:mag} bottom). It is straightforward to deduce the net macroscopic moment and its temperature variation from the latter values, which agree remarkably well with the magnetization curve (Fig. \ref{fig:mag} bottom). In this ferrimagnetic structure, the net moment directed along the $a$ axis ($c$ axis in the parent $I4/mmm$ space group) is simply the difference between the moments on the pyramidal sublattice ($8m$) and the sum of the moments  on the octahedral ($4g$) and tetrahedral ($4g$) sublattices (Fig. \ref{fig:str} and \ref{fig:mag} inset). The octahedral layers CoO$_6$ do not produce any net moment in the proposed model. The magnetization peak observed around 290K is naturally explained by the different critical behaviours of the distinct sublattices (Fig. \ref{fig:mag}) in the CoO$_4$ layer, which seem directly correlated to the number of nearest neighbour magnetic interactions. It appears that the sublattice magnetization has a lower critical exponent (faster saturation) for the octahedral site. The critical behaviours are also affected by the structural phase transition at T$_{S3}$, but in the absence of a detailed model for the crystal structure below this temperature, it is impossible to relate this effect to the displacive modes. Finally, it should be pointed out that in previous models, the presence of a ferromagnetic component due to a canted arrangement required the admixture of at least two irreps, whereas our model involves the single $\Gamma^+_3$ irrep of the $Cmma$ space group in agreement with symmetry requirements for second-order transitions.\\
\indent In conclusion, the complex superstructure in the Sr$_3$YCo$_4$O$_{10.72}$ perovskite is related to the ordering of oxygen vacancies in a pattern forming zigzag stripes in the oxygen-deficient CoO$_{4.72}$ layers. The superstructure is built from a corner-sharing network of Co ions in four-, five- and six-fold coordination, each with distinct spin-states. The magnetic ordering occurring at T$_m\sim$ 360K involves a simple antiferromagnetic arrangement of neighbour spins but the symmetry of the superstructure and the existence of distinct spin states result in a net ferrimagnetic moment. The nature of the oxygen ordering superstructure is key to understand the ferromagnetism, previously ascribed to a canted arrangement. Remarkably, the complex temperature dependence of the magnetization can be quantitatively explained by the different moment values extracted from neutron data. Its peculiar shape can be directly related to the different criticalities of the magnetic order parameters for the Co sublattices. The proposed model can be easily extrapolated for other compositions of the series exhibiting the oxygen-vacancy superstructure but with different $R$-ions and oxygen content. In all cases, the presence of the superstructure results in an effectively different Co coordination in the CoO$_{4+\delta}$ layers and therefore in a resultant uncompensated moment.

\end{document}